\documentclass[a4paper,twocolumn,11pt,unpublished]{quantumarticle}
\pdfoutput=1
\usepackage{amsmath,amssymb,amsfonts,amsthm,bm,braket,mathtools,booktabs}
\usepackage{graphicx}
\usepackage{float}
\usepackage{tikz}
\usetikzlibrary{arrows.meta}
\usepackage{placeins}
\usepackage[hidelinks]{hyperref}
\usepackage{xcolor}

\newcommand{\E}{\mathbb{E}}
\newcommand{\Prob}{\mathbb{P}}
\newcommand{\I}{\mathbb{I}}
\newcommand{\Tr}{\operatorname{Tr}}
\newcommand{\mc}{\mathrm{mc}}
\newcommand{\MCD}{\mathrm{MCD}}

\theoremstyle{plain}
\newtheorem{theorem}{Theorem}

\newtheorem{corollary}[theorem]{Corollary}
\theoremstyle{definition}
\newtheorem{definition}[theorem]{Definition}
\theoremstyle{remark}
\newtheorem*{remark*}{Remark}

\begin{document}

\title{Post-Selection-Free Quantum Automated Learning}

\author{Junkai Wang}
\email{wangtheophys@outlook.com}
\affiliation{Yau Mathematical Sciences Center, Tsinghua University}
\author{Jin-Peng Liu}
\email{liujinpeng@tsinghua.edu.cn}
\affiliation{Yau Mathematical Sciences Center, Tsinghua University}
\affiliation{Institute for Applied Mathematics, Tsinghua University}
\affiliation{Yanqi Lake Beijing Institute of Mathematical Sciences and Applications}

\begin{abstract}
Quantum automated learning (QAL) trains quantum models by directly
updating quantum states, offering an approach to quantum machine
learning that avoids optimizing variational circuit parameters.
However, post-selection constrains the probability of obtaining the
final learned model. We develop a post-selection-free QAL algorithm
that organizes a complete training path into a coherent quantum circuit
and uses fixed-point amplitude amplification to increase the
probability of obtaining the learned model. We provide theoretical
guarantees on output error and learning loss, and identify conditions
under which additional coherent resources reduce the cost of restarting
failed trajectories. The construction gives QAL a coherent
implementation with explicit learning guarantees and resource costs,
enabling a model with certified learning quality and controlled output
error to be obtained with high probability.
\end{abstract}

\maketitle
\section{Introduction}

Quantum automated learning (QAL) trains a quantum state through updates
specified by training examples \cite{Ye2025QAL}. Data-conditioned
contractions suppress incorrect predictions, turning learning into a
state-preparation process. This approach carries the model in a quantum
state and dispenses with variational circuit-parameter optimization.
The original QAL analysis gives learning and survival guarantees under
its spectral assumptions \cite{Ye2025QAL}.

In a measured implementation, each contraction is accepted after a
successful flag measurement. For a fixed order of training examples,
the one-pass success probability is the product of the conditional step
probabilities,
\begin{equation}
P_{\mathrm{chain}}(T)=\prod_{t=1}^{T}p_t.
\label{eq:chain}
\end{equation}
For paths with small survival, measured global restart repeatedly
discards a failed partial trajectory and begins at the initial state.
The cost of obtaining a learned model therefore depends on both its
learning quality and the probability of executing the selected path.

We develop a coherent execution architecture that retains the QAL
learning rule and its projector-mixing updates. An explicit unitary
dilation implements each update with a flag qubit. We leave the flags
unmeasured throughout the predeclared path, then use fixed-point
amplitude amplification (FPAA) to raise the probability of their joint
success branch \cite{Yoder2014FPAA}. The resource paid is additional
coherent execution of the path and its adjoint, together with phase
reflections. The successful branch preserves the original conditional
QAL model exactly. Tracing out all flags and any purification register
returns an unconditional output whose state and learning-loss errors
are controlled by the residual failure target.

A pathwise certificate supplies the information needed to choose that
amplification sequence before execution. It bounds survival and learning
loss jointly for the same sampled order of training examples. The proof
controls an unnormalized loss margin and the survival probability around
an imaginary-time reference state. A low-energy spectral anchor gives
explicit sufficient conditions for a learned model with a positive
survival floor. These conditions determine the amplification degree and
transfer the original QAL generalization guarantee to the returned model.

Probabilistic imaginary-time preparation and monitored quantum dynamics
provide related settings for selected nonunitary evolution
\cite{Motta2020QITE,Nishi2022PITEQAA,Wang2025QSVT}. Existing methods
amplify multistep preparation circuits \cite{Nishi2024MultistepQAA}.
For QAL, our joint certificate links the amplitude to a useful learned
model on the path that the device will execute. The resulting access
model uses coherent preparation of the initial state, reversible
dilation blocks, and reflections about the input and success spaces.

We examine this depth--success trade through learning bounds, exact
circuit calculations, and resource comparisons on the same selected
paths. The calculations check the conditional model, amplified success
probability, and flags-traced output separately. Resource counts include
the prefixes reached before a failed measured restart and the stated
preparation and reflection charges. A separate commuting-projector
family admits a closed-form certificate that identifies a regime with
both low output loss and fewer selected-filter calls. Together, these
results specify how additional coherence improves the probability of
obtaining a QAL model and when it reduces the chosen execution cost.

\section{QAL model and execution protocols}
\label{sec:qal-model}
The register $S$ carries the quantum model, $F$ is a success flag, and $R$
may purify an initially mixed model. A data-encoded unitary $U_t$ and a
correct-label projector $\Pi_t$ specify the projector-mixing update.

\begin{definition}[Flagged update]
A unitary $U_K$ on $F\otimes S$ block-encodes a contraction $K$ when
$\bra{0}_FU_K\ket{0}_F=K$. For a normalized input $\ket{\psi}$, measuring
$F=0$ succeeds with probability
\begin{equation}
p(K,\psi)=\|K\ket{\psi}\|^2.
\label{eq:pdef}
\end{equation}
\end{definition}

At step $t$, set $A_t=U_t^\dagger\Pi_tU_t$ and $H_t=\I-A_t$. Then
\begin{equation}
K_t=(1-\eta_t)\I+\eta_tA_t=\I-\eta_tH_t,
\qquad 0\le\eta_t\le1.
\label{eq:ktdef}
\end{equation}
When a success outcome is selected, the original QAL update is
\begin{equation}
\ket{\psi_t^\star}=K_t\ket{\psi_{t-1}}/
\|K_t\ket{\psi_{t-1}}\|.
\label{eq:idealupdate}
\end{equation}
The normalized update requires a nonzero denominator.

A schedule $\omega=(s_1,\ldots,s_T)$ is drawn independently from one
fixed training distribution before the quantum run. For the selected
filters and initial density operator $\sigma$, write
\begin{align}
K_\omega&=K_{T,s_T}\cdots K_{1,s_1},\nonumber\\
R_\omega&=K_\omega\sigma K_\omega^\dagger,\quad
Z_\omega=\Tr R_\omega,\nonumber\\
\rho_\omega&=R_\omega/Z_\omega\qquad(Z_\omega>0).
\label{eq:fixed-path-state}
\end{align}
The normalized model $\rho_\omega$ is the successful QAL output for
this path. Normalizing the averaged subnormalized state
$\bar R_T=\E_\omega R_\omega$ weights each conditional model by the
declared path-sampling weight times its survival probability.
The path average $\E_\omega\rho_\omega$ uses the declared sampling
weights directly.
For the learning theorem set $\bar H=\E_sH_s$,
$\beta=\sum_t\eta_t$, and $\Gamma=\sum_t\eta_t^2$.
The path theorem permits any fixed sampling distribution. For the
generalization claim, we draw uniformly from an i.i.d. training
set, so $\bar H$ is its empirical loss operator.

Two execution conventions matter. Under \emph{global restart}, a failed
flag discards the partial trajectory and execution begins again at step
one. A \emph{local-repeat} cost assumes the same intermediate input can
be restored or freshly prepared after failure; its preparation cost
must be charged. The local-repeat plots evaluate
success management under this restoration access model.

\section{Learning guarantees and coherent execution}
\label{sec:coherent-qal}

\subsection{Exact update dilation}
The projector-mixing update admits an exact one-flag implementation. This local identity is the starting point for compiling a complete, predeclared path.

\begin{theorem}[Projector-mixing block encoding]
\label{thm:block}
Let $A_t=U_t^\dagger\Pi_tU_t$ with $A_t^2=A_t$ and $0\le\eta_t\le1$, and put $Q_t=\I-A_t$ and
\begin{equation}
K_t=(1-\eta_t)\I+\eta_t A_t=A_t+(1-\eta_t)Q_t.
\label{eq:Kproj}
\end{equation}
For a normalized input $\ket{\psi_{t-1}}$, with
$q_t=\bra{\psi_{t-1}}A_t\ket{\psi_{t-1}}$, the selected flag has probability
\begin{align}
p_t&=\|K_t\ket{\psi_{t-1}}\|^2\nonumber\\
&=(1-\eta_t)^2+\eta_t(2-\eta_t)q_t\nonumber\\
&\ge(1-\eta_t)^2.
\label{eq:ptformula}
\end{align}
Writing $\gamma_t=\sqrt{\eta_t(2-\eta_t)}$, the unitary
\begin{equation}
U_{K_t}=\begin{pmatrix}K_t&-\gamma_tQ_t\\[2pt]
\gamma_tQ_t&K_t\end{pmatrix},
\label{eq:block}
\end{equation}
is an exact selected-branch implementation:
$\bra{0}_F U_{K_t}\ket{0}_F=K_t$.
\end{theorem}

\subsection{Learning quality and survival on one path}
The floor in Eq.~\eqref{eq:ptformula} holds for every normalized input. A useful training path also requires a low-loss model. We certify these two quantities jointly on the \emph{same} sampled path. Let $R_\beta^{\rm ITE}=e^{-\beta\bar H}\sigma e^{-\beta\bar H}$. For $0\le\tau\le1$, take analytic inputs $z_\beta\le\Tr R_\beta^{\rm ITE}$ and $m_\beta(\tau)\le\Tr[(\tau\I-\bar H)R_\beta^{\rm ITE}]$.

\begin{theorem}[Joint path certificate]
\label{thm:traj}
For the predeclared iid path $\omega$ and $u_\zeta=\sqrt{2\Gamma\log(2/\zeta)}$, with probability at least $1-\zeta$,
\begin{align}
Z_\omega&\ge z_\beta-4\Gamma-u_\zeta,\nonumber\\
M_\omega(\tau)&:=\Tr[(\tau\I-\bar H)R_\omega]
\ge m_\beta(\tau)-4\Gamma-u_\zeta.
\label{eq:jointpath}
\end{align}
If both right-hand sides are positive, the same path has a usable survival floor and $\Tr(\bar H\rho_\omega)\le\tau$.
\end{theorem}

The unnormalized margin is essential: concentration applies before dividing by the random $Z_\omega$. The proof combines $\|\E R_\omega-R_\beta^{\rm ITE}\|_1\le4\Gamma$ with one-sample sensitivity at most $2\eta_t$ and is given in Appendix~\ref{sec:supp-v6}. The guarantee is independent of Hilbert-space dimension.

A concrete analytic certificate follows from an initial spectral anchor. Let $w,y>0$, $\widetilde g,\kappa\ge0$, and let $P_{\bar H\le\widetilde g+w}$ denote the corresponding spectral projector. Suppose $\Tr(P_{\bar H\le\widetilde g+w}\sigma)\ge a_0>0$, $\widetilde g\le\kappa w$, and $\widetilde g+2w\le1$. For $\beta=y/w$, put
\begin{equation}
d=a_0e^{-2y(1+\kappa)},\quad
c=1-\frac{(1-a_0)e^{-2y}}{2ea_0y}>0.
\label{eq:anchorconstants}
\end{equation}
When
\begin{equation}
\Gamma\le\min\!\left\{\frac{cwd}{16},
\frac{c^2w^2d^2}{32\log(2/\zeta)}\right\},
\label{eq:anchorbudget}
\end{equation}
Theorem~\ref{thm:traj} gives, on one event of probability at least $1-\zeta$,
\begin{equation}
Z_\omega\ge\frac{3d}{4},\qquad
\Tr(\bar H\rho_\omega)\le\widetilde g+2w.
\label{eq:anchorpath}
\end{equation}
For uniform rates, a sufficient training length is $T=O[y^2\log(2/\zeta)/(c^2d^2w^4)]$.

\subsection{Coherent compilation and amplification}
The selected filters are Hermitian contractions. For any such $K$, let $C_K=\sqrt{\I-K^2}$. Its Julia dilation
\begin{equation}
D(K)=\begin{pmatrix}K&C_K\\ C_K&-K\end{pmatrix}
\label{eq:julia}
\end{equation}
is unitary. Apply the selected dilation at each step with a fresh, unmeasured flag, and let $W_\omega$ denote the resulting circuit. For a purification $\ket{\Psi_0}$ of $\sigma$ and a path with $Z_\omega>0$,
\begin{equation}
W_\omega\ket{0^T}\ket{\Psi_0}
=\sqrt{Z_\omega}\ket{0^T}\ket{\Psi_\omega}
+\sqrt{1-Z_\omega}\ket{B_\omega},
\label{eq:pathbranch}
\end{equation}
where $\ket{B_\omega}$ is orthogonal to the all-zero flag space and the model marginal of $\ket{\Psi_\omega}$ is exactly $\rho_\omega$. Projecting $W_\omega$ onto the all-zero flag subspace yields $K_\omega$. This is the sequential postselection construction of Ref.~\cite{Wang2025QSVT}; the path certificate supplies its learning-dependent amplitude floor.

If $Z_\star\le Z_\omega$ is fixed before circuit execution, fixed-point amplitude amplification preserves the good component and raises its probability to at least $1-\delta$ with
\begin{equation}
Q_{\rm path}=O\!\left(Z_\star^{-1/2}\log(1/\delta)\right)
\label{eq:querylaw}
\end{equation}
calls to $W_\omega$ or $W_\omega^\dagger$ \cite{Yoder2014FPAA}. Each complete-path call uses $T$ selected dilation blocks; preparing and reflecting about the initial state and success space require separate resources. The adjoint call traverses the dilation blocks in reverse order, replacing each block by its unitary adjoint.

\subsection{Returned model and generalization}
\begin{corollary}[Unconditioned learned model]
\label{cor:unconditioned-model}
After amplification, trace out the flags and purification register without accepting any outcome, and call the resulting model $\widetilde\rho_\omega$. For a certified path,
\begin{equation}
\|\widetilde\rho_\omega-\rho_\omega\|_1\le2\delta,
\qquad
\Tr(\bar H\widetilde\rho_\omega)\le\tau+\delta.
\label{eq:unconditioned}
\end{equation}
\end{corollary}

Tracing orthogonal good and bad flag spaces gives $\widetilde\rho_\omega=p_{\rm good}\rho_\omega+(1-p_{\rm good})\xi_\omega$, with $p_{\rm good}\ge1-\delta$. Averaging over paths and charging atypical paths their worst-case loss of one gives mean empirical loss at most $\tau+\delta+\zeta$. For a training set of $N$ independent, identically distributed examples, take the path distribution to be uniform over that set, so $\bar H=N^{-1}\sum_{i=1}^N H_i$. The original QAL generalization event \cite{Ye2025QAL} then gives $\epsilon_{\rm gen}=\sqrt{4\log(2^{n+1}/\delta_{\rm gen})/N}$. On a certified path the population risk of $\rho_\omega$ is at most $\tau+\epsilon_{\rm gen}$, and that of $\widetilde\rho_\omega$ is at most $\tau+\epsilon_{\rm gen}+\delta$. The joint claim holds with probability at least $1-\zeta-\delta_{\rm gen}-\delta_{\rm spec}$ when the spectral premise fails with probability at most $\delta_{\rm spec}$; if the premise is verified only for a selected dataset, the claim is conditional on that dataset.

The purification argument also covers a \emph{prepared} mixed initial state, including a microcanonical double, whenever its success floor and input reflection are available. The preparation circuit and its reflection determine the available access to this mixed input.

\subsection{Extensions to state preparation}
For a projector-mixing step, Eq.~\eqref{eq:ptformula} supplies the
state-independent floor $(1-\eta_t)^2$. A tighter floor may be obtained
from a lower confidence bound on the projector overlap $q_t$, with the
implementation cost including the copies used to estimate that overlap
and any re-preparation of the intermediate model. The
complete-path theorem avoids measuring an unknown intermediate model:
it fixes one classical schedule and certifies its joint amplitude from
the mean loss operator before coherent execution.

The same fixed-point sequence can be applied to other probabilistic
state-preparation primitives when an amplitude floor is known. For an
imaginary-time contraction $K_{\tau}=e^{-\Delta\tau H}$ with
$0\le H\le\Lambda I$, the spectral bound immediately gives
\begin{equation}
p_{\tau}=\bra{\psi} K_{\tau}^{2}\ket{\psi}
\ge e^{-2\Delta\tau\Lambda}=:p_{\star,\tau}.
\label{eq:itefloor}
\end{equation}
Figure~\ref{fig:route2} illustrates this separate imaginary-time
setting. The spectral bound certifies its
state-preparation amplitude. A prepared purification may likewise be amplified as
a pure input when its preparation and reflection are available; this
construction uses the preparation and reflection access associated with
that supplied mixed input \cite{Wang2025QSVT}.

\section{Numerical results}
\label{sec:numerical-results}
\begin{figure}[!t]
  \centering
  \includegraphics[width=\columnwidth]{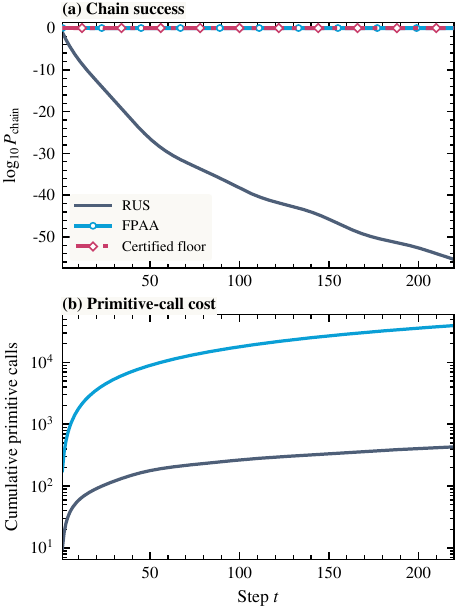}
  \caption{Per-step success and primitive-call costs in a projector-overlap model. The RUS curve in (a) shows single-pass prefix survival; FPAA is shown with its certified floor. Panel~(b) uses local repetition with restoration of each intermediate state after failure.}
\label{fig:route1}
\end{figure}

\begin{figure}[!t]
  \centering
  \includegraphics[width=\columnwidth]{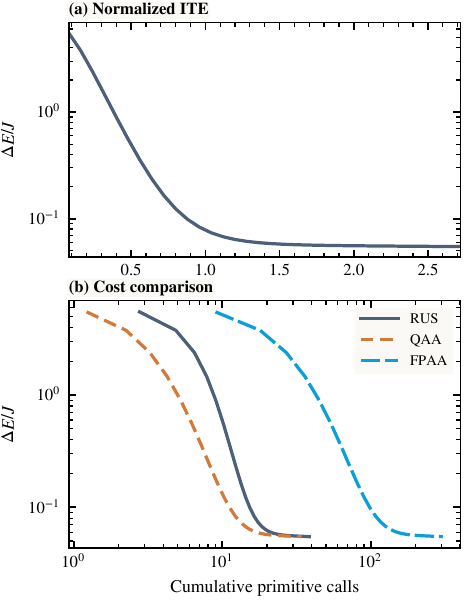}
  \caption{Normalized imaginary-time evolution of a six-site Ising chain. Panel~(a) shows the energy error; (b) plots the same trajectory against schematic primitive-call costs for three success-management methods. Local RUS uses the actual success probability, QAA assumes it is known, and FPAA uses the spectral floor.}
\label{fig:route2}
\end{figure}

Figures~\ref{fig:route1} and \ref{fig:route2} illustrate local success
management for two distinct state-preparation settings. Their cost models
assume local input restoration. We then test the complete-path
construction and a declared per-accepted-output resource model.

The pathwise theorem also has a direct projector-mixing check. For two
noncommuting rank-one sample projectors with analytically known mean
spectrum, its survival and unnormalized-margin bounds select the
complete-path amplification degree before either path is drawn. In
one sampled path, $2{,}500$ updates require degree three and produce an
unconditioned model with training loss $0.448762$ from initial loss
$0.51$; a second certified schedule reaches $0.397924$ with degree
five. Exact small-matrix recurrences evaluate the
ideal-unitary outputs of these paths and their flags-traced channels.
Their direct coherent realization retains thousands of live flags.

\begin{figure}[!t]
\centering
\includegraphics[width=0.98\columnwidth]{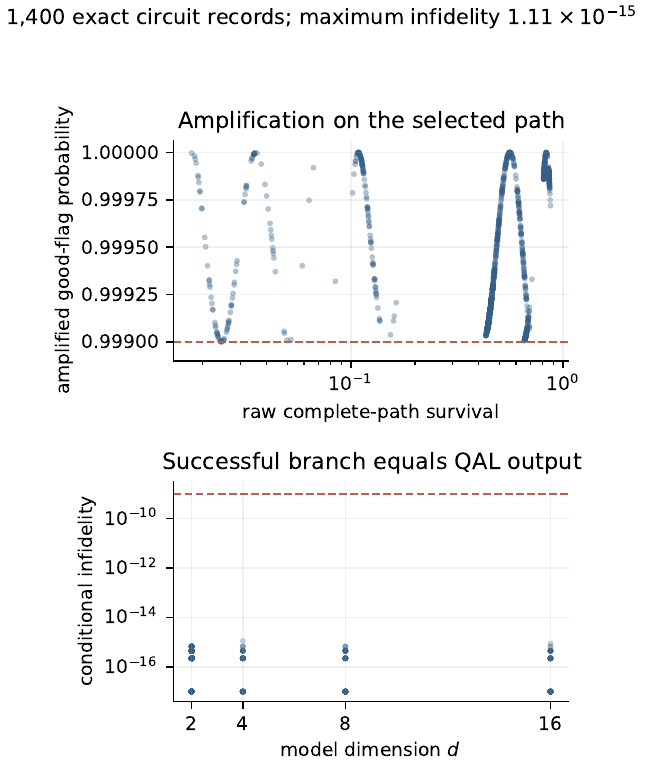}
\caption{Exact circuit checks of amplification and branch preservation. The good-flag probability increases while the successful model retains the conditional QAL output to numerical precision.}
\label{fig:circuit-records}
\end{figure}

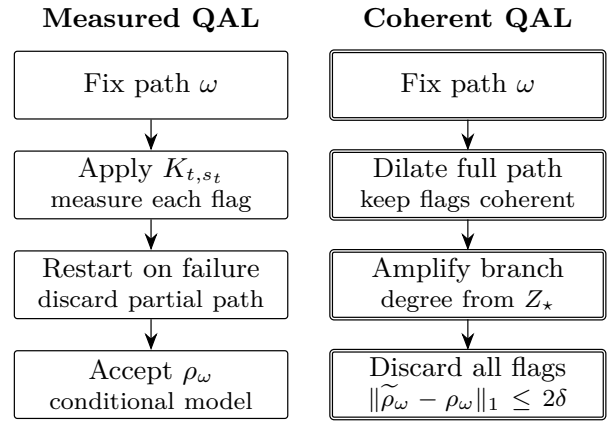
\begin{figure}[!t]
\centering
\resizebox{0.98\columnwidth}{!}{\begin{tikzpicture}[
  >={Stealth[length=2mm]},
  box/.style={draw=black, rounded corners=1pt, fill=white,
    minimum width=3.15cm, minimum height=0.78cm,
    text width=2.95cm, align=center, inner sep=2pt},
  coherent/.style={box, double, double distance=0.45pt},
  every node/.style={font=\footnotesize}
]
\node[font=\footnotesize\bfseries] at (0,0.75) {Measured QAL};
\node[font=\footnotesize\bfseries] at (3.65,0.75) {Coherent QAL};
\node[box] (m1) at (0,0) {Fix path $\omega$};
\node[box] (m2) at (0,-1.15) {Apply $K_{t,s_t}$\\[-1pt]{\scriptsize measure each flag}};
\node[box] (m3) at (0,-2.30) {Restart on failure\\[-1pt]{\scriptsize discard partial path}};
\node[box] (m4) at (0,-3.45) {Accept $\rho_\omega$\\[-1pt]{\scriptsize conditional model}};
\node[coherent] (c1) at (3.65,0) {Fix path $\omega$};
\node[coherent] (c2) at (3.65,-1.15) {Dilate full path\\[-1pt]{\scriptsize keep flags coherent}};
\node[coherent] (c3) at (3.65,-2.30) {Amplify branch\\[-1pt]{\scriptsize degree from $Z_\star$}};
\node[coherent] (c4) at (3.65,-3.45) {Discard all flags\\[-1pt]{\scriptsize $\|\widetilde\rho_\omega-\rho_\omega\|_1\le2\delta$}};
\foreach \a/\b in {m1/m2,m2/m3,m3/m4,c1/c2,c2/c3,c3/c4}
  \draw[->] (\a) -- (\b);
\end{tikzpicture}}
\caption{Two ways to execute a predeclared QAL path. Measured global
restart discards a failed partial path; coherent execution retains the
flags until fixed-point amplification acts on their joint success
space. The good branch has the exact conditional QAL model. Discarding
all flags gives an approximate unconditioned model, under the stated
preparation and reflection access assumptions.}
\label{fig:coherent-flow}
\end{figure}

A separate circuit calculation checks branch preservation in 1,400
finite-dimensional records; another 200 longer records use analytic
amplification formulas (Fig.~\ref{fig:circuit-records}). Under a declared per-accepted-output
scalar cost model, fixed-point amplification is cheaper than measured
restart on 40 of 1,600 heterogeneous configuration records and on none
when counting selected-filter queries alone. These weighted-cost cases evaluate the stated
resource model. The certified low-loss comparison is developed for the
commuting-projector family in Section~\ref{sec:resource-analysis}. Figure~\ref{fig:coherent-flow} contrasts
the measured and coherent executions; Fig.~\ref{fig:cost-records}
shows the cost-model split. Protocols and controls are given in the
Appendix~\ref{sec:supp-v6-numerics}.

\begin{figure}[!t]
\centering
\includegraphics[width=0.98\columnwidth]{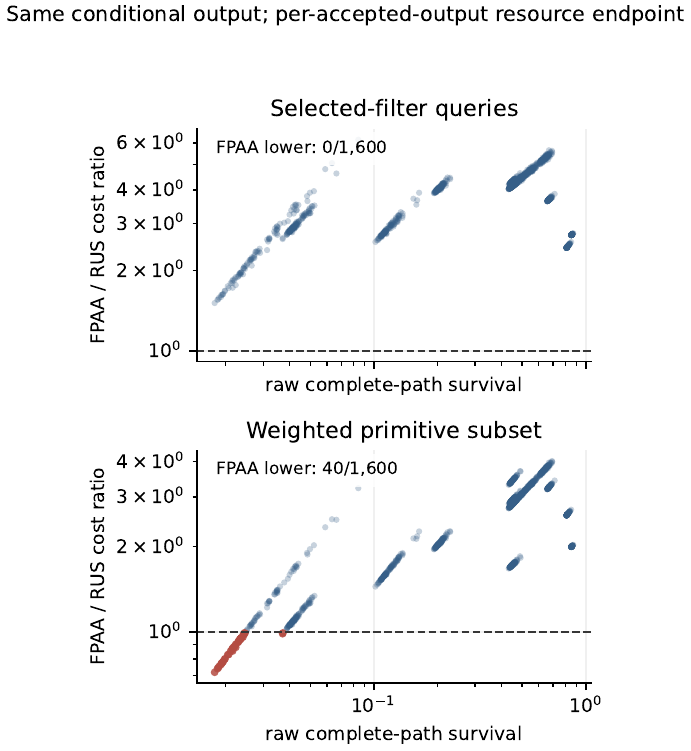}
\caption{Per-accepted-output FPAA-to-global-restart cost ratios. The upper panel counts selected-filter queries; the lower adds weighted preparation, reflection, flag, measurement, reset, and readout charges. Ratios below one occur in the weighted model. Each comparison uses conditional accepted outputs and the specified cost assignments.}
\label{fig:cost-records}
\end{figure}

\section{Resource analysis}
\label{sec:resource-analysis}
The coherent protocol retains the flags across an entire training path
and amplifies their joint success branch. Its depth budget includes
complete-path traversals, adjoint traversals, and phase reflections.
The joint learning-and-survival certificate sets this budget for a path
with a useful model. Tracing out the flags turns the amplified circuit
into a source of unconditional outputs with the error in
Eq.~\eqref{eq:unconditioned}.

The access model includes coherent preparation of the initial model,
reversible selected dilations, and reflections about the input and
success spaces. A local-repeat protocol additionally uses restoration
or re-preparation of the intermediate state after failure. These
operations carry preparation, reflection, and reset costs in the
weighted resource model. Figure~\ref{fig:cost-records} resolves the
resulting comparison between measured global restart and FPAA on a
fixed training path.

\begin{figure}[!t]
  \centering
  \includegraphics[width=\columnwidth]{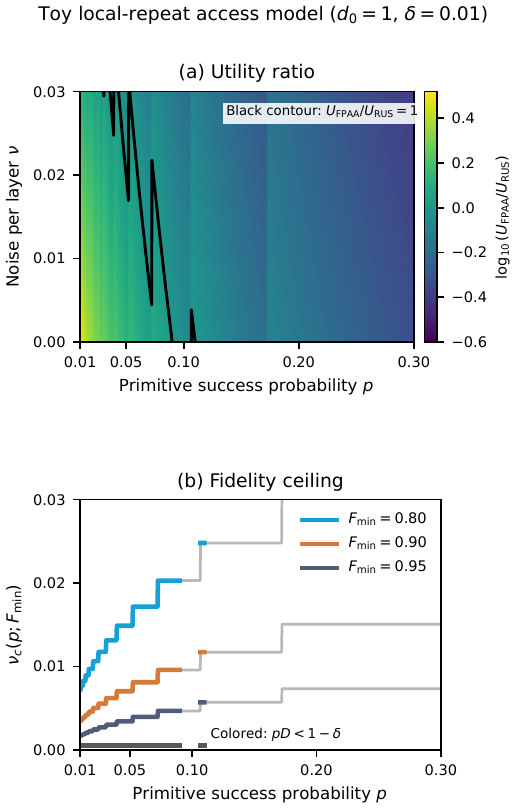}
  \caption{Crossover in a toy local-repeat cost and fidelity model. The heat map shows $\log_{10}(U_{\mathrm{FPAA}}/U_{\mathrm{RUS}})$, with utility $U=F/C$ and a black unity contour. Gray curves mark fidelity ceilings; colored segments also satisfy the local cost condition.}
\label{fig:route3}
\end{figure}

Figure~\ref{fig:route3} evaluates a local-repeat depth-and-fidelity model.
It assigns exponential attenuation to coherent depth and compares the
utility of FPAA with the measured retry protocol. The fidelity ceilings
and cost condition identify where the additional amplification depth
meets a prescribed fidelity target while reducing local retry cost.

\begin{figure}[H]
\centering
\includegraphics[width=0.98\columnwidth]{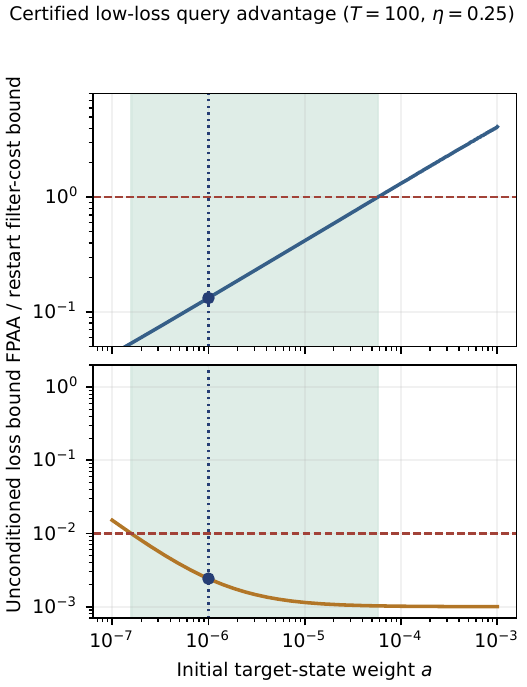}
\caption{Certified low-loss filter-query advantage in a commuting-projector QAL family. The shaded interval simultaneously satisfies the unconditional loss target and a conservative FPAA-to-global-restart cost ratio below one. The marker selects one instance. The ratio counts selected-filter queries alone.}
\label{fig:certified-advantage}
\end{figure}

A four-level commuting-projector family gives an analytic comparison in
which learning quality and filter cost are certified together. On a
predeclared iid path event of probability at least $0.9938$, it has a
survival floor of $10^{-6}$, unconditioned loss below $0.00242$, and an
FPAA-to-global-restart selected-filter cost ratio below $0.133$; see
Fig.~\ref{fig:certified-advantage}. Its common dark state permits a direct
certificate for this learning rate, where the generic bound of
Theorem~\ref{thm:traj} is vacuous. Appendix~\ref{sec:supp-certified-win}
gives the derivation and reproducible path calculation.

The selected-filter ratio counts the learning operations needed to
obtain the conditional model, including partial paths discarded by
measured restart. The flags-traced coherent output has the corresponding
learning-loss guarantee without an acceptance test. Preparation and
reflection charges enter through the overhead condition in
Appendix~\ref{sec:supp-certified-win}. Gate synthesis, data encoding,
readout, and noise determine the additional resources of a device-level
implementation; the filter comparison supplies one component of that
accounting. The shared dark state is also a known basis state, so direct
state preparation gives a distinct resource baseline for this family.

\section{Conclusion}
Projector-mixing QAL can be executed as a coherent complete-path circuit.
Fixed-point amplification raises the probability of its joint success
branch while preserving the learned model on that branch. Discarding
the flags returns a model whose state error and learning-loss penalty
are controlled by the residual failure target.

The joint path certificate connects learning quality to the amplitude
floor that sets the amplification sequence. It gives a post-selection-free
QAL protocol with explicit preparation and reflection access. The
additional coherent depth improves the probability of obtaining the
model, and the commuting-projector family identifies a certified regime
with fewer selected-filter calls than measured global restart. These
results connect the QAL learning guarantee to the resource conditions
for its coherent execution.

\section*{Acknowledgments}
We thank Dong-Ling Deng and Qi Ye for various insightful discussions
and their dedicated effort on the development of Quantum Automated
Learning. J.-P.L. acknowledges support from the Quantum Science and
Technology--National Science and Technology Major Project (Grant
No.~2024ZD0300500), the Excellent Young Scientists Fund Program, and
start-up funding from Tsinghua University and the Beijing Institute of
Mathematical Sciences and Applications.

\FloatBarrier

\clearpage
\onecolumngrid
\appendix
\section{Joint Path Certificate and Complete-Path Compilation}
\label{sec:supp-v6}

This section gives the derivations for the joint survival and loss bound used
in the main text. The schedule is sampled before quantum execution: the $s_t$
are independent draws from one fixed distribution, $0\le H_s\le\I$,
$K_{t,s}=\I-\eta_tH_s$, and $0\le\eta_t\le1$. With
$\bar H=\E_s H_s$, $\beta=\sum_t\eta_t$, and
$\Gamma=\sum_t\eta_t^2$, set
\begin{equation}
R_\beta^{\rm ITE}=e^{-\beta\bar H}\sigma e^{-\beta\bar H},
\qquad \bar R_T=\E_\omega R_\omega .
\end{equation}
The sampled-path state is normalized only after the full sequence of
selected filters.

\subsection{Mean state and one-sample sensitivity}

For positive $X$, compare the averaged filter map
$\Phi_\eta(X)=\E_s[(\I-\eta H_s)X(\I-\eta H_s)]$
with $\mathcal T_\eta(X)=e^{-\eta\bar H}Xe^{-\eta\bar H}$.
Expanding the first map and bounding the exponential remainder using
$0\le H_s,\bar H\le\I$ gives
\begin{equation}
\|\Phi_\eta(X)-\mathcal T_\eta(X)\|_1
\le4\eta^2\Tr X.
\label{eq:supp-v6-onestep}
\end{equation}
Both maps contract trace norm on Hermitian inputs. Telescoping over the
independent draws therefore yields
\begin{equation}
\|\bar R_T-R_\beta^{\rm ITE}\|_1\le4\Gamma.
\label{eq:supp-v6-mean}
\end{equation}

For two schedules differing only at step $t$, let $X$ be their common
positive subnormalized prefix. If $K=\I-\eta_tH$ and
$K'=\I-\eta_tH'$, then
\begin{equation}
KXK-K'XK'=(K-K')XK+K'X(K-K').
\end{equation}
Since $\|K-K'\|_\infty\le\eta_t$ and both filters have norm at most
one, the trace-norm difference at this step is at most
$2\eta_t\Tr X\le2\eta_t$. Later filters are contractive, so
\begin{equation}
\|R_\omega-R_{\omega'}\|_1\le2\eta_t.
\label{eq:supp-v6-sensitivity}
\end{equation}
Trace duality gives the same replacement bound for $Z_\omega$ and
$M_\omega(\tau)=\Tr[(\tau\I-\bar H)R_\omega]$, because
$\|\tau\I-\bar H\|_\infty\le1$ for $0\le\tau\le1$.

For either scalar, the one-sided bounded-differences inequality with
failure probability $\zeta/2$ gives the radius
$u_\zeta=\sqrt{2\Gamma\log(2/\zeta)}$. Equation~\eqref{eq:supp-v6-mean}
and trace duality place their means at least $z_\beta-4\Gamma$
and $m_\beta(\tau)-4\Gamma$, respectively. A union bound proves the
joint inequalities in the main text. Only when the survival lower bound
is positive do we divide by $Z_\omega$; a positive margin then implies
$\Tr(\bar H\rho_\omega)\le\tau$.

\subsection{Anchor-band inputs}

Let $\mu$ be the spectral measure of $\bar H$ induced by $\sigma$.
Under the anchor conditions in the main text, the band
$E\le\widetilde g+w$ contributes at least
\begin{equation}
d_{\rm exact}
=a_0e^{-2y(1+\widetilde g/w)}
\ge d=a_0e^{-2y(1+\kappa)}
\end{equation}
to ideal survival. At $\tau=\widetilde g+2w$, the same band contributes
at least $wd_{\rm exact}$ to the ideal unnormalized margin. For the
negative tail, $x=E-\tau>0$ and
$\sup_{x\ge0}xe^{-2\beta x}=1/(2e\beta)$ give
\begin{equation}
\Tr[(\tau\I-\bar H)R_\beta^{\rm ITE}]
\ge wd_{\rm exact}
\left[1-\frac{(1-a_0)e^{-2y}}{2ea_0y}\right]
\ge cwd.
\label{eq:supp-v6-anchormargin}
\end{equation}
Substituting $z_\beta=d$ and $m_\beta(\tau)=cwd$ in the joint
certificate, the budget in the main text ensures
$4\Gamma\le cwd/4$ and $u_\zeta\le cwd/4$. Thus
$M_\omega(\tau)\ge cwd/2$ and
$Z_\omega\ge d-cwd/2\ge3d/4$. The last inequality uses
$c\le1$ and $w\le1/2$. Uniform rates have
$\Gamma=\beta^2/T$, giving the stated sufficient
$T=O[y^2\log(2/\zeta)/(c^2d^2w^4)]$ scaling.

The width exponent above follows from the sufficient anchor-band
conditions. The scalar example with
independent fair Bernoulli $H_t\in\{0,1\}$ and uniform
$\eta=\beta/T$ has
$Z_\omega=(1-\beta/T)^{2N}$ for
$N\sim{\rm Binomial}(T,1/2)$. Its survival fluctuates on the
$\beta/\sqrt T=\sqrt\Gamma$ scale with nonvanishing probability. This example
identifies the generic fluctuation scale used in the joint concentration
certificate.

\subsection{Branch preservation and output without acceptance}

For a Hermitian contraction $K$, $C=\sqrt{\I-K^2}$ commutes with $K$,
so $D(K)=\left(\begin{smallmatrix}K&C\\C&-K\end{smallmatrix}\right)$
satisfies $D(K)^2=\I$. Applying a fresh dilation flag at each step and
then projecting all flags onto zero gives exactly $K_\omega$. On a
purification of $\sigma$, the normalized all-zero component has model
marginal $\rho_\omega$. Fixed-point amplification acts in the span of
the good and bad components and preserves the direction of the good
component. This proves the exact conditional-state claim.

After the sequence, let $p_{\rm good}\ge1-\delta$ be the all-zero flag
probability. Orthogonality of the flag spaces makes the unconditioned
model
$\widetilde\rho_\omega=p_{\rm good}\rho_\omega+
(1-p_{\rm good})\xi_\omega$ for some state $\xi_\omega$. Hence
$\|\widetilde\rho_\omega-\rho_\omega\|_1\le2\delta$.
For any loss effect $0\le L\le\I$, its expectation can increase by
at most $\delta$. Applied to $L=\bar H$, this gives the stated
$\tau+\delta$ loss. For generalization, sample the path uniformly from
the $N$ i.i.d. training examples, making $\bar H$ their empirical mean.
Combining the joint path event with the original QAL uniform-generalization
event gives the main text's
$1-\zeta-\delta_{\rm gen}-\delta_{\rm spec}$ probability and adds the
inherited square-root generalization term. If the spectral anchor is
asserted only for a selected dataset, this last statement is
conditional on that dataset.

The complete-path query count assumes a reversible unitary dilation
for each selected filter, its adjoint, a preparation circuit for the input
purification, and the required phase reflections. One query traverses
all $T$ selected filters. A local retry uses an independently restored
intermediate input. Its restoration charge enters the cost of executing
the full training trajectory.

\section{Projector-Mixing QAL and the Trajectory Wrapper}
\label{sec:route1supp}

\subsection{Projector algebra and exact dilation}

For one step write $\eta=\eta_t$. Let
\begin{equation}
A_t=U_t^\dagger \Pi_t U_t,\qquad Q_t=\I-A_t .
\label{eq:supp_AQ}
\end{equation}
Because $\Pi_t^2=\Pi_t$ and $U_t$ is unitary, one has $A_t^2=A_t$, $Q_t^2=Q_t$, and $A_tQ_t=Q_tA_t=0$. The projector-mixing update becomes
\begin{equation}
K_t=(1-\eta)\I+\eta A_t=A_t+(1-\eta)Q_t ,
\label{eq:supp_K}
\end{equation}
so that
\begin{equation}
K_t^2=A_t+(1-\eta)^2Q_t,\qquad
\I-K_t^2=\eta(2-\eta)Q_t .
\label{eq:supp_K2}
\end{equation}
Writing $\gamma=\sqrt{\eta(2-\eta)}$, the block matrix
\begin{equation}
U_{K_t}=
\begin{pmatrix}
K_t & -\gamma Q_t\\
\gamma Q_t & K_t
\end{pmatrix}
\label{eq:supp_UK}
\end{equation}
obeys
\begin{equation}
U_{K_t}^\dagger U_{K_t}
=
\begin{pmatrix}
K_t^2+\gamma^2Q_t & 0\\
0 & K_t^2+\gamma^2Q_t
\end{pmatrix}
=\I ,
\label{eq:supp_unitary}
\end{equation}
where Eq.~\eqref{eq:supp_K2} was used in the last equality. The top-left block is therefore exactly $K_t$.

\subsection{Single-step success and structure-aware certification}

For a normalized state $\ket{\psi_{t-1}}$, decompose
\begin{equation}
\ket{\psi_{t-1}}=\ket{\psi_A}+\ket{\psi_Q},
\qquad
\ket{\psi_A}=A_t\ket{\psi_{t-1}},
\qquad
\ket{\psi_Q}=Q_t\ket{\psi_{t-1}} .
\label{eq:supp_decomp}
\end{equation}
Then $\braket{\psi_A|\psi_Q}=0$ and $\|\psi_A\|^2+\|\psi_Q\|^2=1$. Using Eq.~\eqref{eq:supp_K},
\begin{equation}
K_t\ket{\psi_{t-1}}=\ket{\psi_A}+(1-\eta)\ket{\psi_Q},
\label{eq:supp_action}
\end{equation}
hence
\begin{equation}
p_t=\|K_t\ket{\psi_{t-1}}\|^2
=1-\eta(2-\eta)\|\psi_Q\|^2
=(1-\eta)^2+\eta(2-\eta)q_t ,
\label{eq:supp_pt}
\end{equation}
with $q_t=\langle A_t\rangle$. This produces two certification strategies:
\begin{align}
p_{\star,t}^{\mathrm{shot}}
&=
\max\!\left\{0,\hat p_t-\sqrt{\frac{\ln(2/\beta_t)}{2N_t}}\right\},
\label{eq:supp_shot}
\\
p_{\star,t}^{\mathrm{str}}
&=
(1-\eta)^2+\eta(2-\eta)\,q_{\star,t},
\label{eq:supp_structure}
\end{align}
where $\hat p_t$ is the empirical success frequency from $N_t$ trials and $q_{\star,t}$ is a lower confidence bound on $q_t$. The second strategy is preferable whenever the projector observable is experimentally accessible because it injects prior spectral knowledge into the bound.

\subsection{Fixed-point amplification and trajectory failure budget}

The Yoder--Low--Chuang fixed-point polynomial can be written as
\begin{equation}
P_D(\lambda)
=
1-\delta\,
T_D\!\left(
T_{1/D}\!\big(\delta^{-1/2}\big)\sqrt{1-\lambda}
\right)^2 ,
\label{eq:supp_cheb}
\end{equation}
where $D$ is odd and $\lambda$ is the input success probability \cite{Yoder2014FPAA}. If
\begin{equation}
\lambda\ge w_D(\delta):=
1-T_{1/D}\!\big(\delta^{-1/2}\big)^{-2},
\label{eq:supp_passband}
\end{equation}
then $P_D(\lambda)\ge 1-\delta$. A sufficient local degree condition is
\begin{equation}
D_t\arcsin\sqrt{p_{\star,t}}\ge \operatorname{arccosh}(\delta_t^{-1/2}).
\label{eq:supp_degree}
\end{equation}

The trajectory wrapper follows from a union bound. Let $F_t$ denote the event that step $t$ fails either because the certification was wrong or because the amplified flag still failed. Then
\begin{equation}
\Prob\!\left[\bigcup_{t=1}^{T}F_t\right]
\le \sum_{t=1}^{T}\Prob(F_t)
\le \sum_{t=1}^{T}(\beta_t+\delta_t).
\label{eq:supp_union}
\end{equation}
This is a per-step failure budget under the idealized ability to restore
or reprepare each intermediate input. It is distinct from the main text's
complete-path survival-and-loss certificate.

\subsection{Budget allocation}

To expose the optimization principle, write the dominant cost as
\begin{equation}
C_{\mathrm{dom}}(\{\delta_t\})
=
\sum_{t=1}^{T} a_t \log\frac{2}{\delta_t},
\qquad
a_t:=\frac{c_0}{\sqrt{p_{\star,t}}}.
\label{eq:supp_costopt}
\end{equation}
Minimizing Eq.~\eqref{eq:supp_costopt} subject to $\sum_t\delta_t\le \varepsilon$ gives the Lagrangian
\begin{equation}
\mathcal{L}=
\sum_{t=1}^{T} a_t \log\frac{2}{\delta_t}
+\lambda\left(\sum_{t=1}^{T}\delta_t-\varepsilon\right).
\label{eq:supp_lagrangian}
\end{equation}
Stationarity yields
\begin{equation}
-\frac{a_t}{\delta_t}+\lambda=0
\quad\Longrightarrow\quad
\delta_t^\star=\frac{a_t}{\lambda}
=\varepsilon\,\frac{a_t}{\sum_s a_s}
\propto \frac{1}{\sqrt{p_{\star,t}}}.
\label{eq:supp_opt}
\end{equation}
Thus the steps with the smallest certified success floor receive the largest individual failure budget, allocating the global budget across the sequential fixed-point wrapper.

\subsection{Numerical instantiation of the QAL trajectory example}

The local-repeat figure in the main text is constructed as follows:
\begin{enumerate}
  \item An explicit synthetic trajectory $\{q_t\}$ is converted to $\{p_t\}$ via Eq.~\eqref{eq:supp_pt}.
  \item The minimum odd degree satisfying Eq.~\eqref{eq:supp_degree} is then used to evaluate the fixed-point polynomial in Eq.~\eqref{eq:supp_cheb} and to produce the amplified chain-success and cumulative-cost curves.
  \item The same certified floors and degree schedule are used to compare naive retry against fixed-point amplification on the identical physical trajectory.
\end{enumerate}
The trajectory uses synthetic projector overlaps. Its cost curve assumes that each intermediate state can be restored after a failed local attempt.

\section{Imaginary-Time Extension}
\label{sec:route2supp}

\subsection{Normalized imaginary-time steps as post-selected primitives}

Let $H\ge 0$ and define the normalized imaginary-time step
\begin{equation}
K_{\tau}=e^{-\Delta\tau H},
\label{eq:supp_ite_step}
\end{equation}
after the standard spectral shift that sets the ground energy to zero. If the eigenvalues of $H$ lie in $[0,\Lambda]$, then the eigenvalues of $K_{\tau}$ lie in $[e^{-\Delta\tau\Lambda},1]$. Consequently,
\begin{equation}
K_{\tau}^2=e^{-2\Delta\tau H}\succeq e^{-2\Delta\tau\Lambda}\I ,
\label{eq:supp_psd_bound}
\end{equation}
and every state obeys
\begin{equation}
p_{\tau}
=
\bra{\psi}K_{\tau}^2\ket{\psi}
\ge e^{-2\Delta\tau\Lambda}
=:p_{\star,\tau}.
\label{eq:supp_ite_floor}
\end{equation}
Equation~\eqref{eq:supp_ite_floor} is the imaginary-time analogue of the QAL success-floor formula: the Hamiltonian spectrum itself provides the certified passband.

\subsection{Six-site exact-diagonalization example}

The main-text example uses a six-site transverse-field Ising chain
\begin{equation}
H_{\mathrm{TFI}}
=
-J\sum_{i=1}^{N} Z_iZ_{i+1}
-h\sum_{i=1}^{N} X_i ,
\qquad N=6 ,
\label{eq:supp_tfi}
\end{equation}
with periodic boundary conditions. Let $E_0$ be the smallest eigenvalue
of $H_{\mathrm{TFI}}$ and set $H=H_{\mathrm{TFI}}-E_0\I\ge0$.
The script computes exact normalized imaginary-time evolution
\begin{equation}
\ket{\psi_{m}}
=
\frac{e^{-\Delta\tau H}\ket{\psi_{m-1}}}
{\sqrt{\bra{\psi_{m-1}}e^{-2\Delta\tau H}\ket{\psi_{m-1}}}},
\label{eq:supp_exact_evolution}
\end{equation}
from a random initial state. The physical accuracy metrics are
\begin{align}
\Delta E_m &= \bra{\psi_m}H_{\mathrm{TFI}}\ket{\psi_m}-E_0
=\bra{\psi_m}H\ket{\psi_m}, \label{eq:supp_energy_error}\\
F_m &= |\braket{\phi_0|\psi_m}|^2, \label{eq:supp_fidelity}
\end{align}
where $\ket{\phi_0}$ is the exact ground state from diagonalization and $\ket{\psi_0}$ is the random initial state.

\subsection{Comparison cost models}

The state trajectory is identical across the three methods; only the resource assigned to each step changes. We use the schematic costs
\begin{align}
C^{\mathrm{RUS}}_m &= \frac{1}{p_m},
\label{eq:supp_rus_cost}
\\
C^{\mathrm{QAA}}_m &=
\max\!\left\{
1,\,
\frac{\pi}{4\arcsin\sqrt{p_m}}
\right\},
\label{eq:supp_qaa_cost}
\\
C^{\mathrm{FPAA}}_m &=
D_{\tau},
\qquad
D_{\tau}
=
O\!\left(\frac{1}{\sqrt{p_{\star,\tau}}}\log\frac{1}{\delta}\right),
\label{eq:supp_fpaa_cost}
\end{align}
where the QAA line is a continuous oracle-call proxy assuming access to
the true $p_m$, while the FPAA line uses only the spectral floor
$p_{\star,\tau}$ of Eq.~\eqref{eq:supp_ite_floor}. These expressions
assign oracle-call costs to the local success-management step, with
state restoration supplied separately. The Ising figure re-expresses
the same energy trajectory against their cumulative sums.

\subsection{Relation to imaginary-time preparation}

Imaginary-time state preparation supplies a second setting where a spectral bound fixes a success floor \cite{Motta2020QITE}. In this comparison, ordinary QAA is assigned the actual step success probability, whereas FPAA uses only the lower bound; the latter schedule remains valid as the actual probability changes along the trajectory \cite{Nishi2022PITEQAA}. The six-site calculation compares the success-management schedules at the level of these oracle-call proxies.

\section{Mixed-State Boundary and Noise-Aware Crossover}
\label{sec:route3supp}

\subsection{Why mixed-state determinization is different}

For a density matrix $\rho$, the ideal post-selected state is
\begin{equation}
\mathcal{P}_{K}(\rho)=\frac{K\rho K^\dagger}{\operatorname{Tr}(K\rho K^\dagger)} .
\label{eq:supp_postmixed}
\end{equation}
The numerator is linear in $\rho$, but the normalization is not. A universal deterministic CPTP map implementing Eq.~\eqref{eq:supp_postmixed} on arbitrary unknown mixed inputs is therefore unavailable in general \cite{Wang2025QSVT}. A known input purification and its preparation circuit instead support the input-specific construction described below.

\subsection{A minimal noise-aware model}

We use a depth-and-retry model to visualize the crossover. Let $d_0$ be the coherent depth of one primitive attempt and let $\nu$ be an effective noise rate per coherent layer. For RUS and FPAA we assign
\begin{align}
C_{\mathrm{RUS}} &= \frac{d_0}{p}, &
F_{\mathrm{RUS}} &= e^{-\nu d_0}, \label{eq:supp_rus_noise}\\
C_{\mathrm{FPAA}} &= \frac{d_0 D(p,\delta)}{1-\delta}, &
F_{\mathrm{FPAA}} &= e^{-\nu d_0 D(p,\delta)}, \label{eq:supp_fpaa_noise}
\end{align}
where $D(p,\delta)$ is the smallest odd degree satisfying Eq.~\eqref{eq:supp_degree} with $p_\star=p$. The associated utility ratio is
\begin{equation}
\frac{U_{\mathrm{FPAA}}}{U_{\mathrm{RUS}}}
=
\frac{F_{\mathrm{FPAA}}/C_{\mathrm{FPAA}}}{F_{\mathrm{RUS}}/C_{\mathrm{RUS}}}
=
\frac{1-\delta}{pD(p,\delta)}
\,e^{-\nu d_0\left(D(p,\delta)-1\right)} .
\label{eq:supp_utility}
\end{equation}
The corrected ratio crosses unity within the illustrated parameter
window. The heat map therefore includes a unity contour and identifies
where FPAA has greater utility in this local-repeat model. The
calculation assigns fidelity through the exponential depth factors in
Eqs.~\eqref{eq:supp_rus_noise} and \eqref{eq:supp_fpaa_noise} and
compares the utility of successful local outputs.

\subsection{Fidelity-limited feasible region}

For a minimum fidelity $F_{\min}$ for the successful output, FPAA must satisfy
\begin{equation}
e^{-\nu d_0 D(p,\delta)}\ge F_{\min}
\quad\Longleftrightarrow\quad
\nu\le \nu_c(p;F_{\min})
:=\frac{\ln(1/F_{\min})}{d_0D(p,\delta)} .
\label{eq:supp_gammac}
\end{equation}
To beat RUS in wall-clock cost at the same time, one additionally needs
\begin{equation}
\frac{D(p,\delta)}{1-\delta}<\frac{1}{p}.
\label{eq:supp_costwin}
\end{equation}
Figure~\ref{fig:route3}(b) plots the boundary in Eq.~\eqref{eq:supp_gammac} for three fidelity thresholds. Thin gray curves show the full boundaries; colored segments mark points also satisfying Eq.~\eqref{eq:supp_costwin}.

\subsection{Known-purification lifting}
Let $\rho_{\mc}=\Pi_{\mc}/d_{\mc}$ and prepare its purification
$\ket{\MCD}_{RS}=d_{\mc}^{-1/2}\sum_{i\in\mc}\ket{i}_R\ket{i}_S$.
For any Kraus operator $K$ with
$p=\Tr(K\rho_{\mc}K^\dagger)>0$, the normalized vector
$(\I_R\otimes K)\ket{\MCD}/\sqrt p$ purifies
$K\rho_{\mc}K^\dagger/p$. If the preparation and reflection about
$\ket{\MCD}$ are available, pure-state amplitude amplification can
increase the success probability of this particular prepared input.
On the successful branch, tracing out $R$ returns the corresponding
postselected mixed state. Discarding the flags and $R$ gives an
unconditional mixture whose residual failure weight controls the output
error.
The basis-state success probabilities need not be equal. The
construction uses a known purification and its reflection, so its
access model specifies a particular prepared input. The no-go statement
concerns a universal deterministic channel acting on arbitrary unknown
mixed inputs.

\section{Certified QAL Paths and Costs}
\label{sec:supp-v6-numerics}

\subsection{A rank-one-projector learning instance}
For an explicit QAL instance, draw independently from two noncommuting
rank-one loss projectors
\begin{equation}
H_{\pm}=\begin{pmatrix}1/4&\pm\sqrt3/4\\
\pm\sqrt3/4&3/4\end{pmatrix},\qquad
\bar H=\operatorname{diag}(1/4,3/4).
\end{equation}
The initial state is $\sqrt{0.48}\ket0+\sqrt{0.52}\ket1$,
and the target loss is $\tau=0.49$. The known spectrum supplies exact
ideal inputs $z_\beta$ and $m_\beta(\tau)$ to the joint path theorem;
the FPAA degree is fixed before sampling. Table~\ref{tab:v6paths}
reports two realizations whose flags-traced losses meet the target.
The theorem guarantees loss at most $\tau+\delta$ on its certified event.
The paths use Python's \texttt{random.Random} with seeds
\texttt{20261004} and \texttt{20261005}; \texttt{getrandbits(T)}
is read least-significant bit first, with $0$ selecting $H_-$
and $1$ selecting $H_+$.

\refstepcounter{table}\label{tab:v6paths}
\begin{center}
\begin{minipage}{0.9\linewidth}
\small\noindent\textbf{Table~\thetable.} Ideal projector-mixing examples. The first three rows are fixed
before drawing the path. The remaining rows concern one reproducible
realization per schedule. $D$ is the trace distance from the successful
QAL model after all flags are traced. Values are rounded for display;
the amplification degrees use the unrounded certificates.
\begin{center}
\begin{tabular}{@{}lrr@{}}\toprule
 & $\beta=0.5$ & $\beta=0.96$\\\midrule
Updates $T$ & 2,500 & 4,000\\
Certified $Z_\star$ & 0.595 & 0.382\\
Chosen FPAA degree & 3 & 5\\\midrule
Flags-traced loss & 0.449 & 0.398\\
$D$ & 0.00106 & 0.00160\\\bottomrule
\end{tabular}
\end{center}
\end{minipage}
\end{center}

The long paths use exact $2\times2$ recurrences for the good amplitude
and flags-traced channel. A short Julia/YLC circuit separately checks
the success polynomial, branch identity, and flags-traced mixture.
Explicit implementation of the direct construction requires thousands
of coherent flags.

\subsection{Finite circuit and resource calculations}
The exact-circuit subset constructs the complete-path dilation and FPAA
sequence, verifying the successful branch to numerical precision
(Figure~\ref{fig:circuit-records}). The remaining records evaluate
analytic success and cost formulas from path propagation.

For odd degree $L$, the cost charges $(L+1)/2$ forward and $(L-1)/2$
adjoint path-unitary calls; RUS is charged only for prefixes reached
before failure. The declared weighted cost favors FPAA for some records.
RUS selected-filter query counts are at most those of FPAA throughout
these records. The favorable weighted-cost records provide numerical
loss evaluations and resource accounting; joint low-loss certification
is supplied by the commuting family below. Figure~\ref{fig:cost-records}
shows the weighted-cost comparison. The scalar ratio accounts for the declared call weights;
depth, ancillas, output-sample multiplicity, data encoding, and inference
require separate resource accounting.

\subsection{A certified low-loss filter-query winning region}
\label{sec:supp-certified-win}

To certify both the learning and resource conditions
\emph{on the same path}, consider a commuting family with two
equiprobable loss projectors on four levels,
\begin{equation}
H_1=\operatorname{diag}(0,1,1,0),\qquad
H_2=\operatorname{diag}(0,1,0,1),\qquad
\bar H=\operatorname{diag}(0,1,1/2,1/2).
\label{eq:supp-win-projectors}
\end{equation}
Prepare a pure initial state whose computational-basis populations are
$(a,(1-a)/2,(1-a)/4,(1-a)/4)$. Its initial training loss is
$3(1-a)/4$. Draw $T$ samples independently with equal probability,
fix their order before quantum execution, and use
$K_s=\I-\eta H_s$ with $k=1-\eta$. Let $n_s(j)$ count appearances
of sample $s$ in the first $j$ updates, and set $n_s=n_s(T)$.
Commutativity gives the exact path survival and unnormalized loss,
\begin{align}
P_\omega&=a+\frac{1-a}{2}k^{2T}
 +\frac{1-a}{4}\bigl(k^{2n_1}+k^{2n_2}\bigr),
\label{eq:supp-win-survival}\\
\Tr(\bar H R_\omega)&=\frac{1-a}{2}k^{2T}
 +\frac{1-a}{8}\bigl(k^{2n_1}+k^{2n_2}\bigr).
\label{eq:supp-win-loss}
\end{align}
In particular $P_\omega\ge a$ for \emph{every} path, so $a$ fixes
the FPAA sequence in advance.

On the event $\mathcal E_m=\{n_1,n_2\ge m\}$, the two equal-frequency
binomial tails obey
\begin{equation}
\Prob(\mathcal E_m)\ge
1-2\exp\!\left[-2T\left(\frac12-\frac mT\right)^2\right],
\qquad m<T/2.
\label{eq:supp-win-event}
\end{equation}
Since $0<k<1$, Eqs.~\eqref{eq:supp-win-survival} and
\eqref{eq:supp-win-loss} then yield
\begin{align}
P_\omega&\le P_{\max}:=
a+(1-a)\left(\tfrac12 k^{2T}+\tfrac12 k^{2m}\right),
\label{eq:supp-win-pmax}\\
\Tr(\bar H\rho_\omega)&\le
\frac{1-a}{a}\left(\tfrac12 k^{2T}+\tfrac14 k^{2m}\right)
=:L_{\max}.
\label{eq:supp-win-lmax}
\end{align}
Corollary~\ref{cor:unconditioned-model} bounds the flags-traced loss
by $L_{\max}+\delta$ on the same event. This commuting-projector
family uses the direct closed-form certificate above. At the
illustrated learning rate, the $\Gamma=T\eta^2$ term in generic
Theorem~\ref{thm:traj} prevents its lower bounds from certifying
positive survival and loss margin.

For a fixed sampled order, measured \emph{global restart} needs an
expected $C_{\rm RUS}=S_\omega/P_\omega$ selected-filter calls per
complete accepted trajectory, where
$S_\omega=\sum_{j=0}^{T-1}P_j$. At prefix length $j$,
$n_1(j)+n_2(j)=j$; arithmetic--geometric mean gives
$k^{2n_1(j)}+k^{2n_2(j)}\ge2k^j$. Thus, on $\mathcal E_m$,
\begin{align}
S_\omega&\ge S_{\min}:=Ta+
\frac{1-a}{2}\frac{1-k^{2T}}{1-k^2}
+\frac{1-a}{2}\frac{1-k^T}{1-k},
\label{eq:supp-win-smin}\\
C_{\rm RUS}&\ge S_{\min}/P_{\max}.
\label{eq:supp-win-rus}
\end{align}
The smallest sufficient odd fixed-point degree $D$ is chosen from
$a$ and $\delta$, not from the realized $P_\omega$. Charging $DT$
selected-filter calls per amplified run and dividing by the guaranteed
good-branch probability $1-\delta$ gives the conservative
same-conditional-output comparison
\begin{equation}
\frac{C_{\rm FPAA}^{\rm filter}}{C_{\rm RUS}^{\rm filter}}
\le\frac{DT P_{\max}}{(1-\delta)S_{\min}}.
\label{eq:supp-win-ratio}
\end{equation}
The actual postselection-free endpoint costs $DT$ filter calls and
has loss at most $L_{\max}+\delta$; the success-normalized numerator
in Eq.~\eqref{eq:supp-win-ratio} is deliberately conservative because
it prices the exact good branch against RUS's accepted output.

For the schedule in Figure~\ref{fig:certified-advantage}
($T=100$, $\eta=1/4$, $m=33$, $\delta=10^{-3}$), a finite interval of
initial target-state weights simultaneously satisfies the prescribed
output-loss target and the strict cost inequality in
Eq.~\eqref{eq:supp-win-ratio}. Both conclusions hold on the same
certified path event. Fixed-seed path calculations check the bounds,
retaining paths both inside and outside that event.
Equation~\eqref{eq:supp-win-event} supplies the analytic probability bound.

If each complete-path call has an additional preparation-and-reflection
charge $h$ measured in filter-equivalent units, the sufficient cost
inequality remains valid for
\begin{equation}
h<\frac{(1-\delta)S_{\min}}{D P_{\max}}-T.
\label{eq:supp-win-overhead}
\end{equation}
This threshold accounts for preparation and reflection in
filter-equivalent units. The comparison concerns FPAA and measured
global restart within the declared filter-access model. The projectors
commute and share a known computational-basis dark state. Direct
preparation of that state supplies a separate baseline; gate synthesis
and device resources enter their corresponding cost models.

\end{document}